\documentclass[aps,prl,twocolumn,amssymb,superscriptaddress]{revtex4}
\usepackage{subfigure}
\usepackage{graphicx}
\usepackage{rotating} 
\usepackage{color}
 \usepackage{amsmath,amsthm}
\usepackage{epstopdf}
\begin{document}

\title{Dynamical Transition and Heterogeneous Hydration Dynamics in RNA}

\author{Jeseong Yoon}
\affiliation{Korea Institute for Advanced Study, Seoul 130-722, Korea}
\author{Jong-Chin Lin}
\affiliation{Institute for Physical Science and Technology, University of Maryland, College Park, Maryland 20742, USA}
\author{Changbong Hyeon}
\thanks{hyeoncb@kias.re.kr}
\affiliation{Korea Institute for Advanced Study, Seoul 130-722, Korea}
\author{D. Thirumalai}
\thanks{thirum@umd.edu}
\affiliation{Institute for Physical Science and Technology, University of Maryland, College Park, Maryland 20742, USA}

\date{\today}   

\begin{abstract}
Enhanced dynamical fluctuations of RNAs, facilitated by a network of water molecules with strong interactions with RNA, are suspected to be critical in their ability to respond to a variety of cellular signals. Using atomically detailed molecular dynamics simulations at various temperatures of purine (adenine)- and preQ$_1$ sensing riboswitch aptamers, which control gene expression by sensing and binding to metabolites, we show that water molecules in the vicinity of RNAs undergo complex dynamics depending on the local structures of the RNAs. The overall lifetimes of hydrogen bonds (HBs) of surface bound waters are more than at least 1--2 orders of magnitude longer than bulk water.  Slow hydration dynamics, revealed in non-Arrhenius behavior of the relaxation time, arises from high activation barriers to break water hydrogen bonds with a nucleotide and by reduced diffusion of water. The relaxation kinetics at specific locations in the two RNAs show a broad spectrum of time scales reminiscent of glass-like behavior, suggesting that the hydration dynamics is highly heterogeneous. 
Both RNAs undergo dynamic transition at $T = T_D \gtrsim 200$ K as assessed by the mean square fluctuation of hydrogen atoms $\langle x^2\rangle$, which undergoes an abrupt harmonic-to-anharmonic transition at  $T_D$.  The near universal value of $T_D$ found for these RNAs and previously for tRNA is strongly correlated with changes in hydration dynamics as $T$ is altered. 
Hierarchical dynamics of waters associated with the RNA surface, revealed in the motions of distinct classes of water with well-separated time scales, reflects the heterogeneous 
local environment on the molecular surface of RNA.  At low temperatures slow water dynamics predominates over structural transitions. 
Our study demonstrates that the complex interplay of dynamics between water and local environment in the RNA structures could be a key determinant of the functional activities of RNA. 
\end{abstract}

\maketitle

\subsection*{INTRODUCTION}
Water, the most abundant constituent in living organisms,  is an essential determinant of the structure and dynamics, and hence the functions of biomolecules 
\cite{Dickerson1982,Doster1989,Rupley1991,Daniel2003,Roh2006,Caliskan2006,Roh2009}. In the well understood case of proteins the
thermodynamic driving force that minimizes the free energy of proteins arises from the simultaneous requirements to sequester hydrophobic residues from water, and  increase the extent of hydration of the surface-exposed hydrophilic residues. Interactions of hydrophilic residues with water enhance the dynamical fluctuation of proteins, promoting enzymatic activity \cite{Doster1989,Rupley1991,Fitter1996PNAS,Reat2000PNAS,Daniel2003,Roh2006}. 
Dynamic fluctuations of solvated folded proteins, quantified by the motion of hydrogen atoms  using neutron scattering measurement, in aqueous D$_2$O solution exhibit qualitative difference from dry proteins, the latter of which can be experimented in a dry powder form, in glycerol solvent \cite{Tsai00BJ}, or in cryosolvent CD$_3$OD/D$_2$O, DMSO/D$_2$O mixed in various proportions \cite{Reat2000PNAS}.
Temperature dependence of protein fluctuation shows that at low temperature mean square displacement (MSD) of hydrogen atoms $\langle x^2\rangle$ in both wet and dry proteins increases in an identical fashion, but $\langle x^2\rangle$ of wet proteins begins to deviate from its linear temperature dependence at the so called dynamical transition temperature $T_D \approx 200$ K, which coincides with the temperature range where functions of most hydrated proteins begin to emerge \cite{Reat2000PNAS,Ramunssen92Nature}. Although the origin of enhanced amplitude of motion of above $T_D$ continues to be debated, it is clear that there is correlation between enzyme activity and abrupt increase in  $\langle x^2\rangle$ above $T_D$. 

Although not as extensively investigated as proteins, there are a number of studies on the hydration effects on DNA, focusing on hydration-induced conformational change from A- to B-form as well as on the variation in conductivity \cite{Dickerson1982,tao1989BP,denisov1997JMB,Endres2004}. In contrast, much less is known about the nature of water dynamics on RNA molecules and its role in modulating their functions. 
Despite a few early studies \cite{Egli96Biochem,Auffinger1997} the potential similarities and differences in hydration effects between proteins and RNA were first pointed out using a combination of quasi-elastic neutron scattering experiments complemented by preliminary simulations \cite{Caliskan2006,Roh2009}. These and more recent experimental \cite{Chen08PRE,Chu13JPCL} and simulation studies on tRNA \cite{Roh2009} and hairpin ribozyme \cite{Zhang13JPCL} have emphasized the role of relaxation of water localized at the surface of RNA in inducing the dynamical transition at $T_D$.      

There are many reasons to undertake a detailed study of the impact of water on the native state fluctuations of RNA. First, in executing many of their functions RNA accesses low lying excitations, which often might involve local melting of bases as a function of temperature. For example, in prokaryotes there are temperature sensitive elements in the 5$^{\prime}$ untranslated regions (UTRs) which form base pair  with ribosome recognition sites (Shine-Dalgarno sequences) at low temperatures, thus suppressing translation. Recent NMR experiments \cite{Rinnenthal10NAR,Nikolova10RNA} suggest that the melting of these strategically located RNA sequences in response to elevation in temperature, needed for initiating translating, involves potential pre-melting of the hydration shell.  Second, the versatile functional capacity of RNA can be attributed to their ability to access  low free energy conformational   excitations, as demonstrated by probing site specific motions in two regulatory forms of RNAs from HIV-1 \cite{Zhang06Science}. 
By undergoing such motions on time scales on the order of a few nanoseconds RNA molecules adopt a heterogeneous ensemble of conformations, which poises them to recognize and bind a diverse set of ligands. 
The key study on RNAs from HIV-1  shows that the ability to access such states on a relatively short time scale shows the ability of RNA to adapt to changing environmental stimuli, and is encoded in the sequence.  
Thirdly, binding of metabolites to riboswitches to control gene expression is likely linked to local fluctuations in specific regions of co-transcriptionally folded UTR regions of  mRNA. 
In all these examples hydration of RNA is likely to play an important role.

\begin{figure}{}
\includegraphics[width=.5\textwidth]{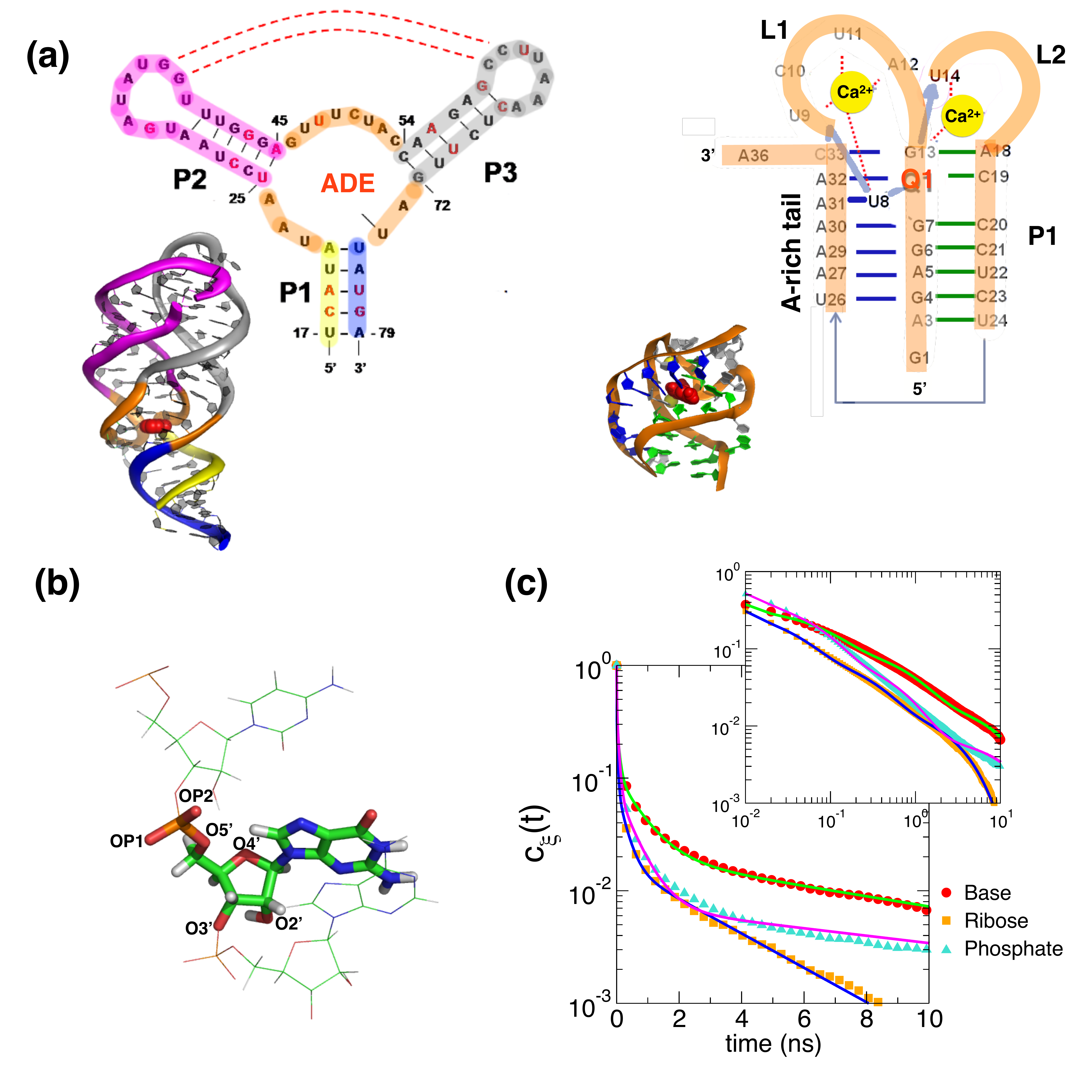}
\caption{(a) Secondary and tertiary structures of purine- (left) and preQ$_1$-riboswitch (right). The tertiary structures  and sizes of both these RNAs, which control gene expression in bacteria by binding adenine (indicated by ADE) and preQ$_1$ (shown as Q1), are dramatically different. The purine riboswitch consists of three paired helices leading to  a three way junction containing the ADE binding region whereas the smaller preQ$_1$ riboswitch forms a pseudoknot. 
(b) Stick representation of a single nucleotide indicating the atomic sites conducive to hydrogen bond formation. 
(c) Correlation functions describing relaxation kinetics of  water HB to the base, ribose, and phosphate groups of the adenine 
riboswitch aptamer ($T=310$ K). 
The data for water HBs with phosphate, ribose, and base are fit using a sum of four exponential functions: 
$c_B(t)=0.673e^{-t/4}+0.219e^{-t/81}+0.086e^{-t/756}+0.022e^{-t/9585}$, 
$c_R(t)=0.689e^{-t/4}+0.228e^{-t/35}+0.066e^{-t/277}+0.017e^{-t/2845}$, 
$c_P(t)=0.569e^{-t/7}+0.352e^{-t/65}+0.071e^{-t/577}+0.008e^{-t/13224}$, where the time constants corresponding to $\tau_i$ are in the unit of ps. The decay of the HB correlations on a log-log scale is in the box.
}\label{fig:nucleic}
\end{figure}

We expect that the unique capacity of a single RNA to change conformations rapidly (plasticity of RNA) should lead to complex solvent dynamics of water \cite{Yoon13JACS,Roh2009,Kuhrova14JCTC} and counterions \cite{WoodsonJMBII01,MisraPNAS01,Chen09JMB,kirmizialtin10JPCB,Kirmizialtin12BJ,Hayes12JACS} that strongly interact with RNA surface.
RNA, which has almost an identical chemical composition with DNA except for an additional hydroxyl group in the 2'-position of ribose ring, displays far more complex structure and dynamics than DNA.  
Unlike double-stranded DNA, characterized with a homogeneous repeat of Watson-Crick base pairs, 
structures of RNA in native state are highly heterogeneous, which could give rise to a great amount of variations in dynamics of water \cite{Caliskan2006,Roh2009} and ions \cite{WoodsonJMBII01,MisraPNAS01,Krasovska06BJ,kirmizialtin10JPCB,Kirmizialtin12BJ,Hayes12JACS} depending on the location on RNA surface.  
In order to elucidate this aspect and to establish the ``universal" structural basis of, particularly, the water-induced dynamical transition in RNA, we focus our study on water dynamics near RNA using two RNA molecules, purine- \cite{Mandal2003Cell,Mandal2003NSMB,Serganov2004,Winkler2005,Cheah2007,Montange2008} and preQ$_1$-riboswitches \cite{Yoon13JACS}. 
Because the water dynamics at the RNA surface is at least 2 orders of magnitude faster than dynamics of monovalent counterions, water dynamics occurs effectively on a static electrostatic environment.
Due to their small size both riboswitch aptamers, containing key structural elements (kissing interactions, non-canonical base pairs, and pseudoknot), are  ideal systems to simulate and to address general features of temperature dependent changes in RNA hydration dynamics. 
Besides usual secondary structural elements such as stack and loop, the purine riboswitch has a three-way junction motif where three helices intersect (Fig.\ref{fig:nucleic}a, left). 
Binding of a purine metabolite into the junction motif consolidates these three helices into a compact structure. 
Similarly, preQ$_1$ ligand and Ca$^{2+}$ ions forge the preQ$_1$-riboswitch into a compact conformation that adopt H-type pseudoknot in which the nucleobases, composing the main helix (P1), stabilize the adenine-rich 3'-tail via Hoogsteen interaction (Fig.\ref{fig:nucleic}a, right) \cite{Zhang11JACS,Yoon13JACS}. 

In this study, we investigate the water dynamics at various locations in the vicinity of folded RNA surface as a function of temperature ($T$) in the range of 100 K $\le T \le$ 310 K  using atomically detailed molecular dynamics simulations. 
To study the relaxation dynamics of water molecules, we computed the number of water hydrogen bonds (HBs) made with each part of RNA structure, and calculated the correlation function to quantify the kinetics and time constant of relaxation ($\tau$) at each temperature. 
To establish the universality of $T_D$, we calculated the $T$-dependence of 
$\langle x^2\rangle$ (see Methods) and showed that RNAs also undergo dynamic transition at $T=T_D\approx 200$ K. 
Over the broad range of temperature above $T=T_D$ the relaxation time of water HB, $\tau$, obeys the Arrhenius relation, which allows us to extract the corresponding activation energy and attempt frequency for the hydrated water around RNA.  
The dynamics of water fluctuations near the riboswitches is highly heterogeneous, reminiscent of glass-like behavior, and is suggestive of low free energy of excitations around the putative folded state.
By comparing these parameters from the Arrhenius-fit with those of bulk water, we show that there are distinct classes of water molecules (not all waters are the same!) near RNA surface.  The heterogeneous behavior of RNA dressed with water molecules, with dynamics in the nanosecond time scale, might have functional importance. 
  
\subsection*{RESULTS \& DISCUSSIONS}
{\bf Population of water and water hydrogen bond on the surface of RNA:}
Before addressing the hydration dynamics on RNA, we calculated the average number of water molecules coordinating the riboswitch aptamers after equilibrating the entire system composing of RNA and solvent at $T=310$ K.
We find that within 3.5 \AA\ from the surface (non-hydrogen heavy atoms) of the purine riboswitch there are on average N$_{H_2O}=744$ water molecules (water oxygens), which corresponds to n$_{H_2O}=10.5$ waters per nucleotide. 
For preQ$_1$-riboswitch, N$_{H_2O}=$361 and n$_{H_2O}=$10.0. 
These water molecules can be defined as those in the first hydration shell around the aptamers. We find that in our simulations that the hydration level, $h$, defined as the amount of hydrated water in gram per 
1 g of RNA, is $h\approx 0.59$ for the purine riboswitch, and $h\approx$0.56 for preQ$_1$ riboswitch.
The relation $h_{\mathrm{purine}}>h_{\mathrm{preQ_1}}$ indicates that preQ$_1$ riboswitch is more compact than purine riboswitch.  
The calculated values of $h$ are similar to the experimentally measured value for hydrated tRNA ($h\approx 0.61$) \cite{Roh2009}. Thus, generic aspects of hydration dynamics can be inferred from our simulations. 

To reveal dynamic features of water near the RNA surface in further detail, we counted the average number of water HBs formed with the riboswitch aptamers. Although there are many ways of defining hydrogen bonds \cite{Kumar07JCP}, we use a geometric criterion, which is often used to analyze hydrogen bonds in proteins and RNA structures. 
We assume that a hydrogen-bond is formed when, in addition to the distance criterion ($<3.5$ \AA), the angle between the hydrogen donor-acceptor axis and the donor-hydrogen bond (O-H/N-H) is $< 30^o$. 
With this definition, there are N$_{HB}\approx$ 543 HBs formed between water molecules and the purine-riboswitch during the simulations, which corresponds to n$_{HB}\approx $ 7.7 water HBs per nucleotide. For preQ$_1$-riboswitch, we find N$_{HB}\approx$ 252 and n$_{HB}\approx$ 7.0. Our previous simulations of tRNA$^{Phe}$ \cite{Roh2009}, whose size is comparable to \textit{add}-riboswitch, showed a similar value for the number.    

We find that the formation of HB with RNA is heterogeneous. In particular, the propensity to make H-bond with water varies depending on the position of a given atom in the nucleotide.  
In the native state, 
(i) the O1P and O2P atoms of phosphate group (Fig. \ref{fig:nucleic}b), which are the most exposed parts of the nucleotide to the solvent, form about $2-2.5$ hydrogen
bonds with water molecules. 
(ii) In most of the residues, atoms O3', O4', and O5' in the ribose group are less solvent accessible, and form only about 0.5, 0.5, and 0.3 HBs, respectively,  whereas O2' atoms form $\approx 1$ HB with water hydrogens (Fig.\ref{fig:nucleic}b).
(iii) In the base groups, unprotonated nitrogen and oxygen (N/O) atoms form about 0.5-1 HB with water hydrogens, 
but the protonated nitrogen and oxygen (NH/OH) form $< 0.5$ HBs with water oxygen. 
As  expected, the average number of HBs with water for protonated sites are smaller than unprotonated sites 
because many protonated sites participate in the formation of base pairs.
\\

{\bf Kinetics of water hydrogen bond near RNA surface:} 
The abundance of water molecules around phosphate group is easily anticipated because phosphate groups are not only negatively charged, but are also the most exposed to the solvent environment in both the native and unfolded states. 
Thus, it may be tempting to conclude that the residence time of water molecules near phosphate groups should be longer than ribose or base groups. 
However, our simulations show that water dynamics around RNA is considerably more complicated reflecting complex structure of RNA.
Surprisingly, it turns out that instead of phosphate groups, water exhibits the slowest relaxation dynamics near bases.  
To elucidate the complex water dynamics near the surface of RNA, we quantify the relaxation dynamics of the number of water HBs as a function of time using the time correlation function ($c(t)$) (see Methods).  

\begin{figure*}{}
\begin{center}
\centerline{\includegraphics[width=0.90\textwidth]{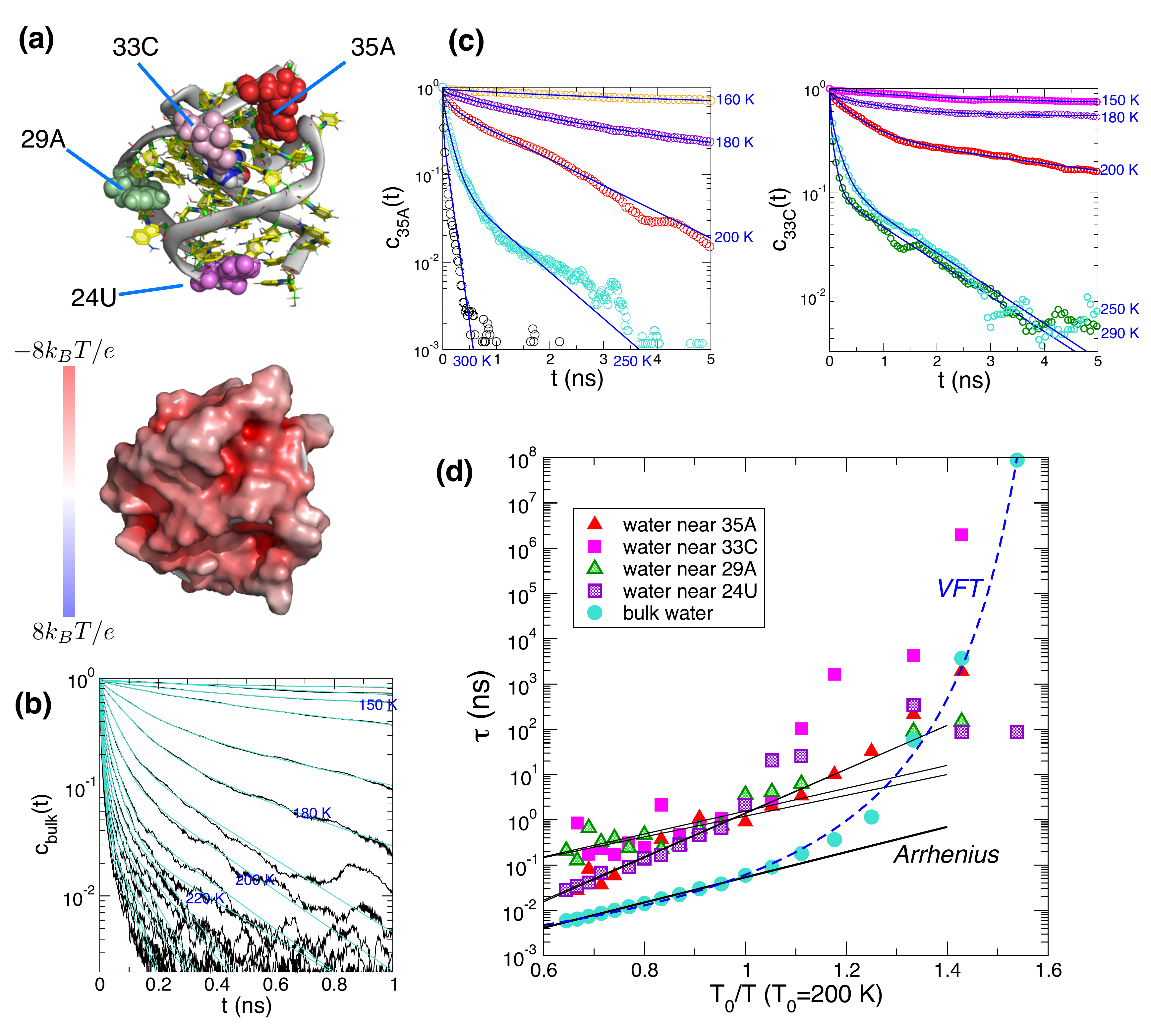}}
\caption{Relaxation kinetics of water HB around RNA surface and in the bulk. 
(a) The structures of preQ$_1$-riboswitch is shown on the top, and the corresponding electrostatic potential at the molecular surface, obtained by numerically solving the nonlinear Poisson Boltzmann equation with the APBS package \cite{Baker01PNAS}, is shown below.
(b) Temperature dependence of the relaxation kinetics of water H-bond in the bulk is shown for reference. 
(c) Same as (b) except the left panel shows the HB relaxation kinetics at the nucleotides 35A (left) and 33C (right). 
Multi-exponential functions (blue lines) are used to fit the data at varying temperatures: 
$c^{35A}_{160K}(t)=0.05e^{-t/0.76 \text{ ps}}+0.20e^{-t/2887 \text{ ps}}+0.85e^{-t/41976 \text{ ps}}$,
$c^{35A}_{180K}(t)=0.09e^{-t/0.61 \text{ ps}}+0.30e^{-t/932 \text{ ps}}+0.61e^{-t5133 \text{ ps}}$,
$c^{35A}_{200K}(t)=0.12e^{-t/0.60 \text{ ps}}+0.28e^{-t/128 \text{ ps}}+0.60e^{-t/1441 \text{ ps}}$,
$c^{35A}_{250K}(t)=0.22e^{-t/0.15 \text{ ps}}+0.63e^{-t/102 \text{ ps}}+0.15e^{-t/592 \text{ ps}}$,
$c^{35A}_{300K}(t)=0.26e^{-t/0.14 \text{ ps}}+0.60e^{-t/20.7 \text{ ps}}+0.14e^{-t/108.5 \text{ ps}}$ for 35A; 
$c^{33C}_{150K}(t)=0.02e^{-t/0.59 \text{ ps}}+0.26e^{-t/2127 \text{ ps}}+0.72e^{-t/5.96\times 10^6 \text{ ps}}$,
$c^{33C}_{180K}(t)=0.14e^{-t/42.5 \text{ ps}}+0.30e^{-t/1048 \text{ ps}}+0.56e^{-t/181483 \text{ ps}}$,
$c^{33C}_{200K}(t)=0.11e^{-t/13.7 \text{ ps}}+0.54e^{-t/533 \text{ ps}}+0.35e^{-t/6469 \text{ ps}}$,
$c^{33C}_{250K}(t)=0.41e^{-t/20.7 \text{ ps}}+0.46e^{-t/163 \text{ ps}}+0.13e^{-t/1277 \text{ ps}}$,
$c^{33C}_{290K}(t)=0.48e^{-t/13.4 \text{ ps}}+0.42e^{-t/85.0 \text{ ps}}+0.10e^{-t/1299 \text{ ps}}$ for 33C. 
\noindent (c) Temperature dependence of water HB relaxation times calculated for the water molecules near 33C, 35A of preQ$_1$-riboswitch, and for the bulk water. 
The onset of deviation from Arrhenius-like behavior ($\tau^{\text{Arr}}\sim e^{A/T}$) is displayed as temperature ($T$) decreases below $T=200$ K.    
The linear regression using $\log{\tau^{\text{Arr}}}=\log{\tau_o}+A/T$, made for the data points with $T\geq 200$ K, i.e., $T_0/T\leq 1$, results in
$(A_{\text{bulk}},\tau^{\text{bulk}}_o)=(4.14,0.088)$, 
$(A_{24U},\tau^{24U}_o)=(7.24,0.018)$,
$(A_{29A},\tau^{29A}_o)=(3.41,6.11)$,
$(A_{33C},\tau^{33C}_o)=(3.75,4.65)$,
and $(A_{35A},\tau^{35A}_o)=(7.17,0.021)$, 
where $A$ has the unit of $k_BT_r$ with $k_B$ being the Boltzmann constant and $T_r=310$ K, and $\tau_o$ has the ps unit. }
\label{fig:water_preQ1}
\end{center}
\end{figure*}

As shown in Fig.\ref{fig:nucleic}c, the relaxation kinetics of water HB at $T=310$ K with different nucleotide groups (B:base, R:ribose, P:phosphate) of purine riboswitch aptamer can be described using a multi-exponential function 
$c_{\xi}(t)=\sum_{i=1}^N\phi_ie^{-t/\tau_i}$ with $\sum_{i=1}^N\phi_i=1$, $\xi=$B, R, and P.  (It is possible to fit $c(t)$ using stretched exponential function. However, we find that depending on $T$ the fit might require more than one stretched exponential function, making the interpretation difficult.)
The multi-phasic kinetics implies that various types of interactions with distinct time scales govern the dynamics of water molecules with nucleotides.  
In Fig.\ref{fig:nucleic}c, up to the 99.9 \% of the decay of $c_{\xi}(t)$ can be described with N=4. 
Among the four phases, the time scales of relaxation and the corresponding weights satisfy an inequality $\tau_1<\tau_2<\tau_3<\tau_4$ with $\phi_1>\phi_2>\phi_3>\phi_4$. 
The values of $\phi_i$, satisfying $\phi_1+\phi_2>0.9$ and $\tau_1<\tau_2<100$ ps, indicate that most of water H-bond dynamics ($\approx$ 90 \%) occur on the time scale of $\sim(1-100)$ ps. 
The correlation function (or survival probability) of water H-bonds drops below $10\%$ when time $t$ is greater than $\sim$ 100 ps. 
At short times ($t < 100$ ps), hydrogen bonding of phosphate groups with water has a longer relaxation time than that of ribose or base with water, 
but at longer times ($t>100$ ps) slow water dynamics is dominated by water molecules trapped near base and ribose although such contribution is made by less than 10 \% of the hydrated water population. 
From $\langle\tau_{\xi}\rangle=\int^{\infty}_0 c_{\xi}(t)dt$, the average lifetime of water H-bond at each group is calculated as     
$\langle\tau_P\rangle=163$ ps, $\langle\tau_R\rangle=81$ ps, and $\langle\tau_B\rangle=289$ ps.  
It is worth emphasizing that the lifetime of water H-bond with base group is twice longer than that with ribose or phosphate group. 
Compared with the lifetime of HBs in bulk water ($\sim 5$ ps) at 310 K (see Fig.\ref{fig:water_preQ1}c), 
the overall relaxation time of hydrogen bonds between nucleotides and water is $1-2$ orders of magnitude longer, which suggests that the slow dynamics of water near RNA is due to stronger interactions with the nucleotide or reduced diffusion constant of water on the RNA surface.  
Although counterions also play a central role in modulating RNA structure \cite{kirmizialtin10JPCB,Kirmizialtin12BJ,Hayes12JACS} and binding or release dynamics of Na$^+$ ions at the surface of RNA would certainly perturb the water environment,  the time scales of such events are far slower than the time scale of hydration. 
The relaxation time of counterion dynamics at the surface of RNA is $\langle\tau_B\rangle\approx 9.2$ ns, $\langle\tau_R\rangle\approx 62.6$ ns, and $\langle\tau_P\rangle\approx 294$ ns, which are more than two orders of magnitude slower than that of water molecules, the details of which we will report elsewhere. 
Due to the large time scale separation between water and ion dynamics, the hydration dynamics around RNA is expected to occur effectively in a static quenched counterion environment. 

The hierarchy of time scales needed to describe the slow hydration dynamics is reminiscent of glassy dynamics. As stated above, it is also possible to fit the relaxation data using stretched exponential. 
Nevertheless, the multi-phasic kinetics, originating from the heterogeneity of conformations that RNA samples on fast time time scales facilitated by water (acting as a lubricant), should be distinguished from glass-like behavior in highly supercooled materials at low temperatures (see below, Fig.\ref{fig:water_preQ1}). 
At $T$=310 K, we show that the majority ($\gtrsim$90 \%) of water HB dynamics are well described by considering double exponential function with two distinct time scales. 
Just like proteins \cite{Williams94ProtSci,Tarek00BJ}, water molecules interacting with RNA exist in diverse forms, which leads to a broad temporal scale in dynamics. 
Based on the well-separated  time scales describing the decay of $c(t)$ we classify water molecules into 
bulk ($\sim \mathcal{O}(10^0-10^1)$  ps), 
surface ($\sim \mathcal{O}(10^1-10^2)$  ps), 
cleft ($\sim \mathcal{O}(10^2-10^3)$  ps), 
and 
buried water ($\gg \mathcal{O}(10^3)$  ps) \cite{denisov1997JPCB} groups. 
\\

{\bf Hydrogen bond dynamics varies in different RNA regions:}  
For adenine riboswitch we performed a similar analysis of HB relaxation dynamics by focusing on the paired helices P1, P2, P3 and the junction region (Fig.\ref{fig:nucleic}a left).
The relaxation kinetics of hydrogen bond at $T=310$ K can be quantitatively fit using,
\begin{align}
c_{P1}(t)&=0.65e^{-t/4.8}+0.29e^{-t/42.3}+0.06e^{-t/532}\nonumber\\
c_{P2}(t)&=0.61e^{-t/4.9}+0.30e^{-t/52.8}+0.09e^{-t/403}\nonumber\\
c_{P3}(t)&=0.62e^{-t/4.8}+0.29e^{-t/52.1}+0.09e^{-t/456}\nonumber\\
c_{J}(t)&=0.57e^{-t/5.4}+0.29e^{-t/65.5}+0.14e^{-t/932}
\label{eqn:fit2}
\end{align}
where $t$ is in the unit of ps, which gives 
$\langle\tau_{P1}\rangle=88.6$ ps, $\langle\tau_{P2}\rangle=96.3$ ps, $\langle\tau_{P3}\rangle=88.7$ ps, and $\langle\tau_{J}\rangle=416$ ps. 
Longer lifetime of water H-bond kinetics in the junction region ($\langle\tau_{J}\rangle>\langle\tau_{P1}\rangle$, $\langle\tau_{P2}\rangle$, $\langle\tau_{P3}\rangle$) implies a greater population of water molecules trapped around the junction region, which is stabilized by the three helices through binding of the purine metabolite.  
Due to the stronger electrostatic potential and the extent of water molecules being buried at the junction where the three helices meet, the dynamics of waters at the junction are less dynamic than on the paired helices.
\\

\begin{figure}{}
\begin{center}
\centerline{\includegraphics[width=.5\textwidth]{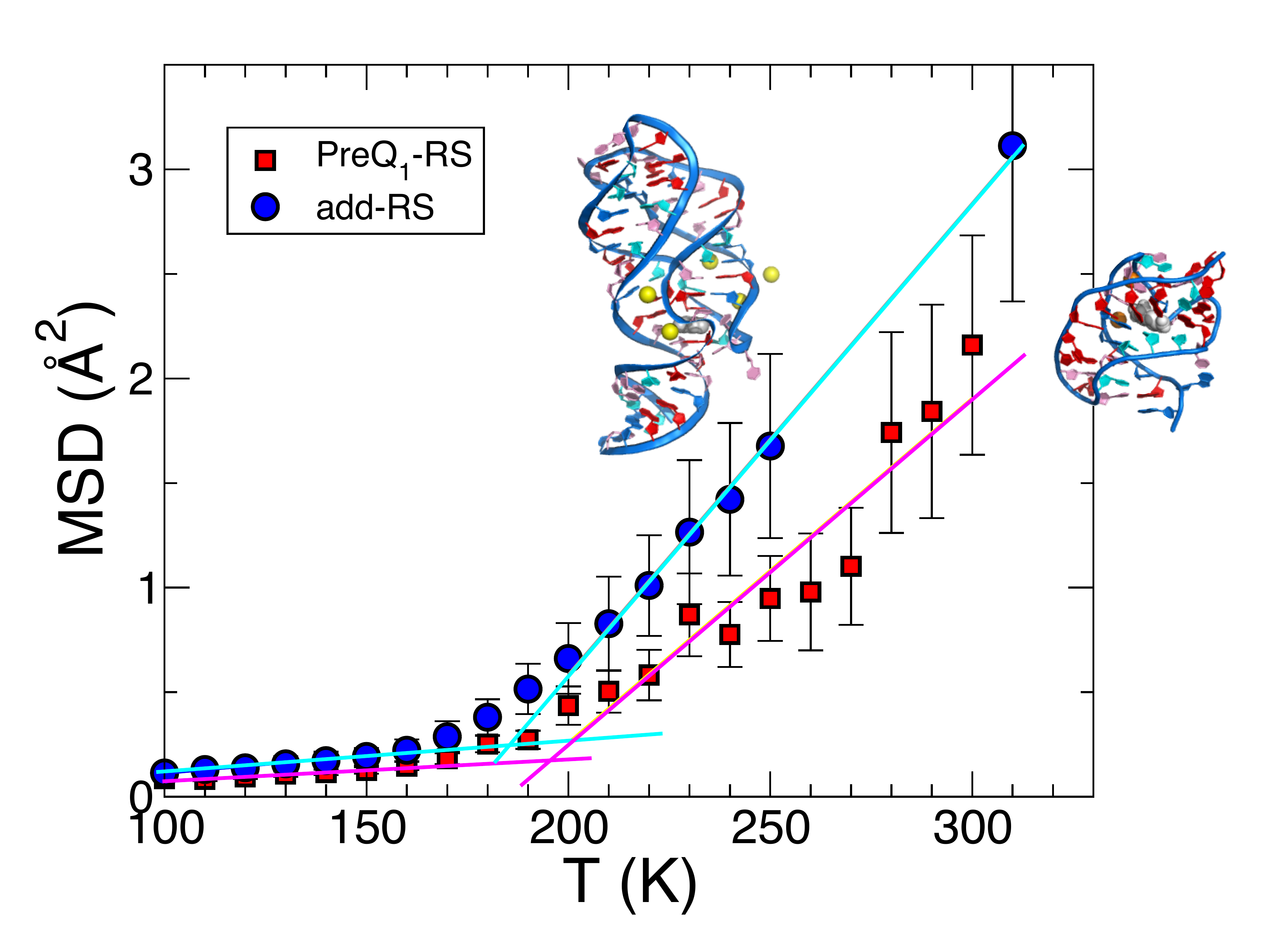}}
\caption{Dynamical transition in riboswitches. Mean square displacement of hydrogen atoms $\langle x^2\rangle$ (defined in Eq.\ref{eqn:x2}) calculated using the time interval of $\delta t=0.1$ ns as a function of temperature.  Dynamical transition temperature of the two riboswitches are in the temperature range of $T=180 - 200$ K. The larger amplitude of motion for the purine riboswitch above $T_D$ relative to preQ$_1$ is related the sizes of the RNA. 
}\label{fig:msd_RS}
\end{center}
\end{figure}

{\bf Dynamical Transition of RNA at around 200 K: } 
A number of studies have confirmed the existence of dynamic transition of proteins at $T_D\approx 180-220$ K, 
above which the temperature response of $\langle x^2\rangle$ of proteins changes from harmonic to anharmonic \cite{Fitter1996PNAS,oostenbrink04JCC,Reat2000PNAS}. Similarly, starting with our early studies it is now known that RNA molecules also undergo a similar transition at $T_D$. It is likely that the transition in RNA is induced by the solvent \cite{Caliskan2006,Zhang13JPCL} although there are differences in the dependence of $T_D$ on the extent of hydration ($h$). In order to establish the universality of the previous findings, established for tRNA,
we calculated MSD of hydrogen atom using the time scale of 0.1 ns for both purine- and preQ$_1$ riboswitches. 
In accord with the previous study using t-RNA \cite{Caliskan2006}, our simulations also confirm that the dynamical transition at $T\approx T_D$ (Fig.\ref{fig:msd_RS}). 
At low temperature range ($T\lesssim 180$ K), MSD for both riboswitches increase almost linearly  with the identical amplitude, but there is an onset of deviation between the two MSDs at $T\approx 190$ K. 
The amplitude of MSD for purine riboswitch (71 nt) is larger than that of preQ$_1$ riboswitch (36 nt), which is consistent with an expectation of $\langle x^2\rangle\sim N$. 

The harmonic-to-anharmonic transition of MSD in response to an increasing temperature is reminiscent of the interpretation for proteins that the folded protein samples alternate conformations in the rugged free energy landscape  
\cite{Fenimore02PNAS}. In the case of RNAs, it is well known that their folding landscape is rough implying that there are easily accessible (on the time scale of nanoseconds) excitations around the native state \cite{Thirum05Biochem}. Thus, it is likely that the mobility of the water molecules in the vicinity of the RNA surface at $T > T_D$ enables structural relaxation \cite{Auffinger02BiophysChem}. 
Recent experimental studies using dielectric and two-dimensional infrared spectroscopies that directly probe both the hydration and protein dynamics provide evidence that protein dynamics is slaved to the surrounding solvent environment \cite{king2012JACS,Khodadadi2011JPCB}.
For proteins, the structural interpretation of the change in dynamics at $T_D$ has been controversial with some ascribing it to the rotational motion of methyl group \cite{lee2001Nature}, while others argue that it is due to solvent dynamics. 
In contrast to the dynamics of proteins, the situation in RNA is simpler. Nucleic acids do not contain methyl groups, and hence it does not seem plausible that the change in the dynamics at $T_D$ is mediated solely by structural transition in RNA. It is likely that the dynamical transition RNA is   ``induced" by a change in water dynamics \cite{Caliskan2006}.
The hypothesis that the dynamical transition of both protein and RNA is water-induced is also supported by the observation that the H-bond relaxation time of bulk water undergoes dynamic transition at $T\approx 200$ K, which is comparable to the dynamic transition temperature of surface water (Fig.\ref{fig:water_preQ1}c, see below).   
\\

\begin{figure*}{}
\begin{center}
\centerline{\includegraphics[width=.8\textwidth]{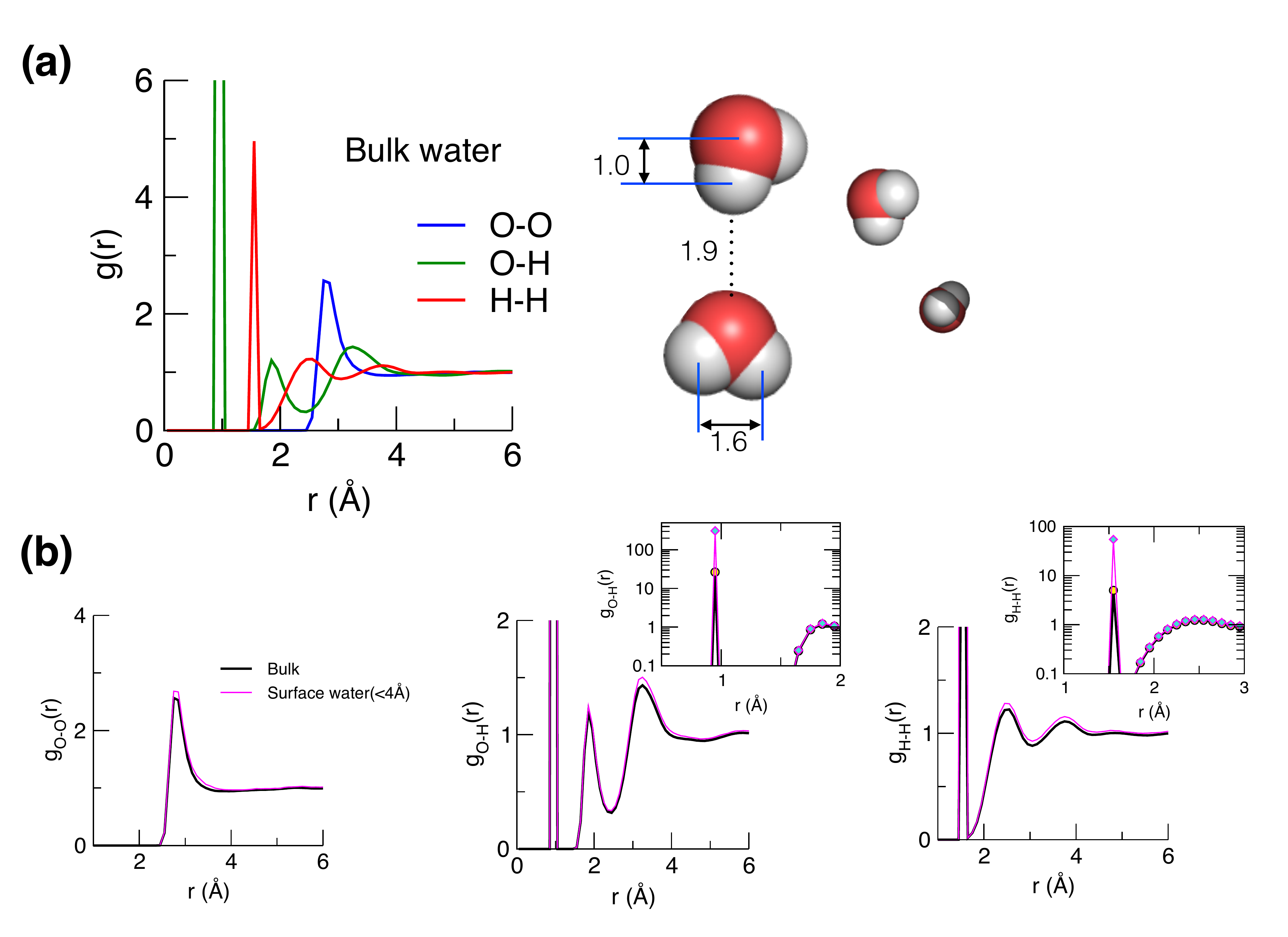}}
\caption{Enhanced ordering of surface water. 
(a) Three radial distribution function (RDF) of bulk water molecules. The first peaks of $g_{O-H}(r)$ ($r\approx 1.0$ \AA) and $g_{H-H}(r)$ ($r\approx 1.6$ \AA) result from the geometry within water molecule. 
(b) RDFs of water molecules within 4 \AA\ from the RNA surface in reference to those at bulk calculated in (a).   
}\label{fig:water_H_bond}
\end{center}
\end{figure*}

{\bf Enhanced ordering of surface water: }
Even at $T=310$ K, compared with radial distribution function (RDF) of bulk water, the first peaks of surface water O-H and H-H RDFs display significantly sharpened peaks (Fig.\ref{fig:water_H_bond}),  
suggesting that there is an enhanced ordering in the surface water. 
The first peaks of surface water RDF, corresponding to the \emph{intramolecular} bond and geometry of water (O-H and H-H), have an order of magnitude greater amplitude than those of bulk water (Fig.\ref{fig:water_H_bond}b), whereas the first peak of O-O pair distribution, which is due to purely \emph{intermolecular} interactions with bulk water molecules, has almost no difference.  
The enhanced ordering of surface water is due to stronger interaction with specific atoms on the RNA surface resulting in  suppression of the  fluctuations of water molecule directly coordinated to the nucleotides. This static picture of water RDF with sharpened peak, is indicative of the enhanced ordering of surface water molecule, also manifests itself  in the water dynamics near the surface (see below).\\

{\bf Heterogeneity of water dynamics on RNA surface: }
Electrostatic potential (Fig.\ref{fig:water_preQ1}a) calculated on the solvent accessible RNA surface reveals that charge balance on RNA surface is heterogeneous, which should result in the dynamics of hydrated water varying depending on the local structure of RNA.  Using the native structure of preQ$_1$-riboswitch aptamer and the electrostatic potential calculated by numerically solving nonlinear Poisson Boltzmann equation in 150 mM-monovalent salt condition as a reference (Fig.\ref{fig:water_preQ1}a), we probe the water dynamics around 24U, 29A, 33C and 35A. 
In 29A and 33C, the phosphate groups, fully exposed to the solvent environment, have a small electrostatic potential at the water-contacting surface due to screening by counterions. The base groups are hidden deep inside the neighboring base-stacks, and are not easily accessible to the solvent. 
In 24U and 35A, the base groups are accessible to the solvent although the base group of 35A is located in the interior of the pocket formed with the L1 loop \cite{Yoon13JACS}, and hence has high negative charge (Fig.\ref{fig:water_preQ1}a).  By studying the relaxation dynamics around these nucleotides with differing local environment, we establish that water dynamics is heterogeneous. 

Fig.\ref{fig:water_preQ1}b shows the correlation function ($c(t)$) of water HB in the bulk (left) and around 35A (right) at various temperatures. 
Similar to the $c(t)$ calculated for B, R, P group of nucleotides at $T=310 K$, 
both $c(t)$ in the bulk and near the nucleotides 35A and 33C are well described by multi-phasic kinetics, i.e., $c(t)=\sum_{i=1}^N\phi_ie^{-t/\tau_i}$. 
For water hydrogen-bond both in the bulk and near 35A and 33C, at high temperature, population of fast dynamics is dominant, but as temperature decreases the influence of slow dynamics grows (see Appendix and Fig.\ref{fig:weight_tau} for the bulk water and the caption of Fig.\ref{fig:water_preQ1} for the waters near 35A and 33C).
 
{\it Dynamics of bulk water:} 
Aside from the water dynamics near the surface of biopolymers, dynamical behavior of bulk water as a function of  temperature is also  a complex topic. 
Thus, it would be useful to discuss the bulk water property prior to discussing the behavior surface water.   
The correlation times of HB for \emph{supercooled} water in the bulk are described either with multi-phasic kinetics up to $\gtrsim 90$ \% of the population (the left panel on the Fig.\ref{fig:water_preQ1}b) or conventionally with stretch exponentials \cite{Gallo1996PRL}. The time scale for the slow relaxation mode reaches $\mathcal{O}(10^{11})$ ps at $T=130$ K. 
The water dynamics, governed by short time dynamics ($\tau\sim \mathcal{O}(1)$ ps), is most dominant at high temperature, $T\approx 300$ K (see APPENDIX), but the anomalies of water dynamics, characterized by long relaxation times, are further amplified as temperature is decreased. 
This observation signifies that motions of water is governed by very different types of phenomenology depending on the time scale of observation, such as short-time free diffusion, intermediate dynamics due to cage effect, followed by long-time free diffusion.
It is noteworthy that even at low temperatures the contribution of fast relaxation mode ($\tau\sim\mathcal{O}(1)$ ps), which could be considered as the bulk water property at high temperatures, does not completely vanish but still remains finite at relatively long times.

As long as $T\gtrsim 200$ K, when the average relaxation time of water HB ($\tau$) is calculated using $\tau=\int^{\infty}_0dtc(t)$, the dependence of $\tau$ on the inverse temperature 
($1/T$) can be described using ``Arrhenius-like" kinetics (Fig.\ref{fig:water_preQ1}c). 
The linear regression using $\log{\tau^{\text{Arr}}}=\log{\tau^{\text{bulk}}_o}+A/T$, made for the data points with $T\geq 200$ K, i.e., $T_0/T\leq 1$, results in
$A_{\text{bulk}}=4.14$ $k_BT_r$ with $T_r=310$ and $\tau^{\text{bulk}}_o=0.088$ ps. 
Of particular note is that the value of $\tau^{bulk}_o$ is similar to Eyring's transition theory estimate $h/k_BT_r\approx 0.15$ ps  \cite{EyringJCP35}. 
The HB relaxation time of supercooled water can be described over the entire temperature range ($150 \leq T\leq 310$ K) 
by using the Vogel-Fulcher-Tamman (VFT) equation often used to analyze temperature dependence of shear viscosity in glass forming materials. 
When the $\tau$ vs $T$ plot for bulk water is fitted to Vogel-Fulcher-Tammann (VFT) equation, $\tau^{\text{VFT}}=\tau^{\text{VFT}}_oe^{\frac{DT_c}{T-T_c}}$ (dashed line in Fig.\ref{fig:water_preQ1}c), we obtain $D=3.00$, $T_c=116.3 K$, and $\tau^{VFT}_o=0.95$ ps. The small value of $D$ suggests that bulk water may be a strong liquid at low temperatures.

{\it Dynamics of surface water:} 
Compared with bulk water, the dynamics of water molecules near RNA is slower by 1-2 orders of magnitude in the studied temperature range (Figs.\ref{fig:water_preQ1}b, c).
Water molecules hydrated near RNA surface, however, also have a hierarchical structure in dynamical property (Fig.\ref{fig:water_preQ1}).   
At $T=300$ K, the water HB dynamics is well fit using $c(t)=0.26e^{-t/0.14ps}+0.60e^{-t/20.7ps}+0.14e^{-t/109ps}$, suggesting that the water HB dynamics is governed by the time scale of $\mathcal{O}(10^1)-\mathcal{O}(10^2)$ ps with the contribution from bulk water-like property being only $\approx 26$ \%.  
It would be more appropriate to interpret the hydrated water near RNA surface with longer time scale by using, RNA structure-dependent description such as, ``surface", ``cleft", and ``buried"-water instead of ``cage effect" in supercooled water in the bulk. 
As temperature is decreased, the contribution of the ``bulk"-water ($\tau\sim\mathcal{O}(1)$ ps) diminishes and the contributions from ``surface", ``cleft", and ``buried"-water becomes more dominant in the population of water interacting with RNA.  

The relaxation dynamics of HB for surface water can also be analyzed using Arrhenius equation over the temperature range $T>180$ K (Fig.\ref{fig:water_preQ1}c): 
(i) $A_{24U}, A_{35A}\approx 2A_{29A}, 2A_{33C}, 2A_{bulk}$ suggests that the activation energies associated with breaking water H-bond from 24U and 35A are twice larger than those associated with 29A, 33C or bulk waters; 
(ii) Each class of $\tau^{\text{bulk}}_o$, $\tau^{24U}_o$, $\tau^{35A}_o$ $(\approx (0.01-0.08)$ ps) and $\tau^{29A}_o$, $\tau^{33C}_o$ ($\approx (4-6)$ ps) has the time constant in a similar range. 
Note that 24U and 35A have a base group accessible to the solvent environment, while 29A and 33C do not.  
The similarity of time constants of 24U and 35A with that of the bulk water indicates that the diffusional property of water interacting with 24U and 35A are effectively identical to that of bulk water.   
Therefore, the above analysis allows us to divide the water dynamics near RNA at least into two classes. 
One is the class in which activation barrier for water HB disruption is increased but the time constant for vibration remains identical to bulk water (water around 24U, 35A); the other is the class in which activation barrier remains identical but the time constant for vibration is reduced by two orders of magnitude from that of bulk water (water around 29A, 33C).

\subsection*{CONCLUDING REMARKS}
Despite fundamental difference between the chemical composition of  RNA and proteins (for example RNA with limited chemical diversity in the nucleotides is highly charged along the phosphodiester backbone whereas there is considerable diversity in the makeup of natural protein sequences), 
we find that globally there are some similarities in the  hydration dynamics in these biopolymers. 
Similar to proteins, RNA displays a dynamical transition signaling a harmonic-to-anharmonic change in the MSD at the near universal dynamical temperature, $T_D \approx 200$ K. This near universal value for $T_D$ suggests that the dynamical transition has more to do with solvent dynamics than with any conformational fluctuations associated with RNA or protein, as this study and others previously have surmised \cite{Tournier03BiophysJ,Caliskan2006}.  

Water molecules interacting with RNA surface exhibit a broad spectrum of time scales, reflecting the underlying architecture of the folded RNA. 
At the ambient temperature, dynamics of bulk water occurs on the time scale of $\sim\mathcal{O}(1)$ ps as assessed in terms of hydrogen bond formation and disruption \cite{Luzar96PRL}.  In sharp contrast,
near RNA surface the time scale of water HB with nucleotides exceeds $\sim\mathcal{O}(10^2)$ ps. 
Although bulk and surface water molecules constitute the major population near RNA surface, there is non-negligible population of water molecules whose time scale of interaction with nucleotide  extends beyond nsec. 
From the simulation trajectories we find that these water molecules are either trapped in the narrow space formed between base-base stack, coordinated with multiple phosphate oxygens simultaneously or with specifically bound multivalent counterions \cite{Allner2012JCTC}. Fluctuations in these regions, which clearly require coordinated motion of several water molecules, must play an important role in ligand recognition by RNA, as suggested by NMR experiments \cite{Zhang06Science}. We classify these as functional water molecules because the dynamics associated with these discrete waters might be responsible for the motions identified in experiments \cite{Rinnenthal10NAR,Nikolova10RNA}. 

Our focus here was restricted to the elucidation of very slow dynamics associated with surface waters and those discrete molecules in the cleft of RNA, which remains folded. An intriguing question is: what role do  these water molecules play in the folding process? 
It is known that the persistence length of RNA decreases dramatically as RNA folds \cite{Caliskan05PRL,Roh11JACS}. This implies that the ordered state of RNA is more flexible than the unfolded  state, which is drastically different compared to proteins. Based on the present study, we speculate that in the unfolded state, with exposed phosphate and bases, the dynamics of water associated with RNA is also slow, perhaps even more so than when RNA is folded. In this case, we expect that water molecules have to disorder to some extent before ions can stabilize the folded state. Thus, in the folding of RNA water molecules should exhibit rich dynamics. 

The dynamics of nucleic acids spans far greater time scales than proteins. Consequently, over a period of time RNA samples a larger class of conformations with ease because of the small stability gap \cite{Thirum05Biochem} separating the folded states and alternate structures. As a result it is likely that RNA, even in the functional state, exhibits persistent heterogeneity or molecule-to-molecule variations \cite{AlHashimi08COSB,Solomatin10Nature,Hyeon2012NatureChem}.  
It is plausible that in conjunction with ion dynamics, temporal scale spanned by water in association with RNA also stretches out over many orders of magnitude.   If this were the case then discrete water molecules whose motions are coordinated with RNA structural transitions might be involved in function as well.

Structural motions lubricated by the discrete water molecules associated with RNA are pivotal in facilitating  molecular interactions  with cosolvents as well as with ligands or metabolites.  
Recently, we  showed that water molecules are  critical for ``inducing" the base pair fluctuations, so that urea molecule can interact more easily with nucleobases \cite{Yoon13JACS}. In the absence of surface water, the urea concentration required for denaturing nucleic acids rises to as high as 10 M \cite{Herskovits63Biochemistry}. Similarly, dehydration of water around RNA due to stabilizing osmolytes such as TMAO involves interplay of interactions between water and cosolvents \cite{Denning13BiophysChem}.  These aspects are worthy of future investigations.

\subsection*{METHODS}

{\bf All-atom Molecular Dynamics (MD) Simulation: }
We used the molecular dynamics package NAMD \cite{Philips2005} to perform atomically detailed simulations for the
71-nt adenine riboswitch (PDB code: 1Y26) \cite{Serganov2004} and 36-nt preQ$_1$ riboswitch (PDB code: 2L1V) \cite{Yoon13JACS} aptamers under CHARMM27 force field.
For adenine riboswitch containing five bound Mg$^{2+}$ ions, we  
added 60 sodium ions by placing each ion around the phosphate group of
RNA backbone, to make the whole system charge neutral. The system is then 
solvated using the SOLVATE program in the VMD package \cite{Humphrey1996} in an
explicit TIP3P water solvent box. A buffer of water are added around the molecule for at least 15 \AA\ in all directions, resulting in total 63,632 atoms in the system. 
PreQ$_1$ riboswitch is solvated in a 60 \AA\ $\times$ 60 \AA\ $\times$ 60 \AA\ box containing 6267 TIP3P water. 
To neutralize the charges on the phosphates and the  two Ca$^{2+}$-bound to the aptamer we placed 51 Na$^+$ and 20 Cl$^-$ ions, which results in $\sim$ 150 mM salt concentration in the bulk, randomly in the box. The resulting system was equilibrated as described elsewhere  \cite{Yoon13JACS}.  
Non-bonded interactions are smoothly switched to zero between 10 and 12 \AA\, 
yielding a cutoff radius of 12 \AA. 
The systems are periodically replicated.  
The particle-mesh Ewald algorithm is used for treating long-range electrostatic 
interactions with a grid spacing smaller than 1 \AA\ \cite{Darden1993}.
The integration time step is 2 fs with the SHAKE method being used
\cite{Ryckaert1977}. The energy of the system is first minimized and gradually heated to the desired temperature. 
We generated a $5-10$ ns production run  at each temperature under
constant $N$, $P$, and $T$ conditions.

Despite inaccuracies (especially for divalent cations such as Mg$^{2+}$) in the current force field for nucleic acids \cite{Allner2012JCTC,Chen13PNAS}, our simulation and analysis on hydration dynamics show semi-quantitative agreements with experimental measurements. Our conclusion on water dynamics near RNA surface as well as at the bulk still holds.
\\

{\bf Mean square displacement or fluctuation of hydrogen atoms: }
The operational definition of $\langle x^2\rangle$ is, 
\begin{align}
\langle x^2\rangle=\frac{1}{N_H}\frac{1}{T-\delta t}\int^{T-\delta t}_0\sum_{i=1}^{N_H}\left[\vec{x}_i(t+\delta t)-\vec{x}_i(t)\right]^2dt
\label{eqn:x2}
\end{align} 
where $\vec{x}_i$ is the position of $i$-th hydrogen atom, $N_H$ is the total number of hydrogen atoms in a molecule, $T$ is the length of trajectory, and $\delta t$ is the time interval used to compute the MSD. \\

{\bf Correlation function to probe water hydrogen-bond kinetics: }
To probe the dynamics of water HB at the surface of RNA quantitatively, 
we calculate the correlation function defined as follows \cite{Chandler1996}
\begin{align}
c(t) &= \langle h(0)h(t)\rangle\nonumber\\
&=\frac{1}{N}\frac{1}{T-t}\int^{T-t}_0\sum_{i=1}^Nh_i(t'+t)h_i(t')dt'.
\label{eqn:corr}
\end{align}
The function $h(t)$ describes whether a HB is formed between
an atom in RNA and a water molecule at time $t$. 
If a HB is formed, the value of $h$ is 1, otherwise, $h = 0$.
The symbol $\langle\ldots\rangle$, which is explicitly written in the second line, denotes a time average along the trajectory and an ensemble average over the number of HBs ($N$) we are interested in.  
The correlation function $c(t)$ corresponds to the survival probability of the HB at time
$t$, given that it is intact at time 0.
\\

{\bf Acknowledgements: }
C.H. thanks Korea Institute for Advanced Study for providing computation resources. D.T. acknowledges a grant from the National Science Foundation (CHE 09-10433).
 
 \begin{widetext}
\section*{APPENDIX}
For completeness we analyzed the relaxation kinetics of hydrogen bonds of water in the bulk. Tri-exponential function fits at various temperatures to the hydrogen-bond dynamics of bulk water in Fig.\ref{fig:water_preQ1}b. 
In the correlation functions below the time is in ps unit. 

$c^{bulk}_{310K}(t)=0.614e^{-t/1.62}+0.357e^{-t/7.99}+0.039e^{-t/55.4}$, 

$c^{bulk}_{300K}(t)=0.563e^{-t/1.74}+0.395e^{-t/7.99}+0.042e^{-t/54.6}$, 

$c^{bulk}_{290K}(t)=0.580e^{-t/2.03}+0.379e^{-t/9.54}+0.041e^{-t/63.4}$, 

$c^{bulk}_{280K}(t)=0.508e^{-t/2.15}+0.432e^{-t/9.10}+0.060e^{-t/59.46}$, 

$c^{bulk}_{270K}(t)=0.597e^{-t/2.76}+0.374e^{-t/13.72}+0.029e^{-t/109.6}$, 

$c^{bulk}_{260K}(t)=0.542e^{-t/2.90}+0.407e^{-t/13.79}+0.051e^{-t/94.52}$, 

$c^{bulk}_{250K}(t)=0.450e^{-t/3.08}+0.482e^{-t/13.49}+0.068e^{-t/93.56}$, 

$c^{bulk}_{240K}(t)=0.482e^{-t/4.07}+0.462e^{-t/18.76}+0.056e^{-t/131.5}$,

$c^{bulk}_{230K}(t)=0.367e^{-t/4.20}+0.548e^{-t/18.69}+0.085e^{-t/123.0}$, 

$c^{bulk}_{220K}(t)=0.547e^{-t/7.62}+0.413e^{-t/37.05}+0.040e^{-t/264.9}$, 

$c^{bulk}_{210K}(t)=0.351e^{-t/7.18}+0.574e^{-t/33.38}+0.075e^{-t/221.0}$, 

$c^{bulk}_{200K}(t)=0.264e^{-t/7.43}+0.592e^{-t/40.032}+0.144e^{-t/235.0}$, 

$c^{bulk}_{190K}(t)=0.115e^{-t/3.90}+0.667e^{-t/49.69}+0.218e^{-t/259.8}$,

$c^{bulk}_{180K}(t)=0.093e^{-t/3.23}+0.659e^{-t/97.83}+0.248e^{-t/463.4}$,

$c^{bulk}_{170K}(t)=0.066e^{-t/2.47}+0.521e^{-t/129.75}+0.413e^{-t/720.4}$,

$c^{bulk}_{160K}(t)=0.066e^{-t/6.35}+0.243e^{-t/287.9}+0.691e^{-t/1571}$,

$c^{bulk}_{150K}(t)=0.052e^{-t/1.45}+0.411e^{-t/543.4}+0.537e^{-t/1.07\times 10^5}$,

$c^{bulk}_{140K}(t)=0.033e^{-t/0.953}+0.756e^{-t/2307}+0.211e^{-t/1.74\times 10^7}$,

$c^{bulk}_{130K}(t)=0.049e^{-t/9.53}+0.139e^{-t/453.2}+0.812e^{-t/1.09\times 10^{11}}$. 
\end{widetext}
\clearpage 

\setcounter{figure}{0}
\makeatletter 
\renewcommand{\thefigure}{A\@arabic\c@figure}
\makeatother
\begin{figure}{}
\begin{center}
\centerline{\includegraphics[width=0.50\textwidth]{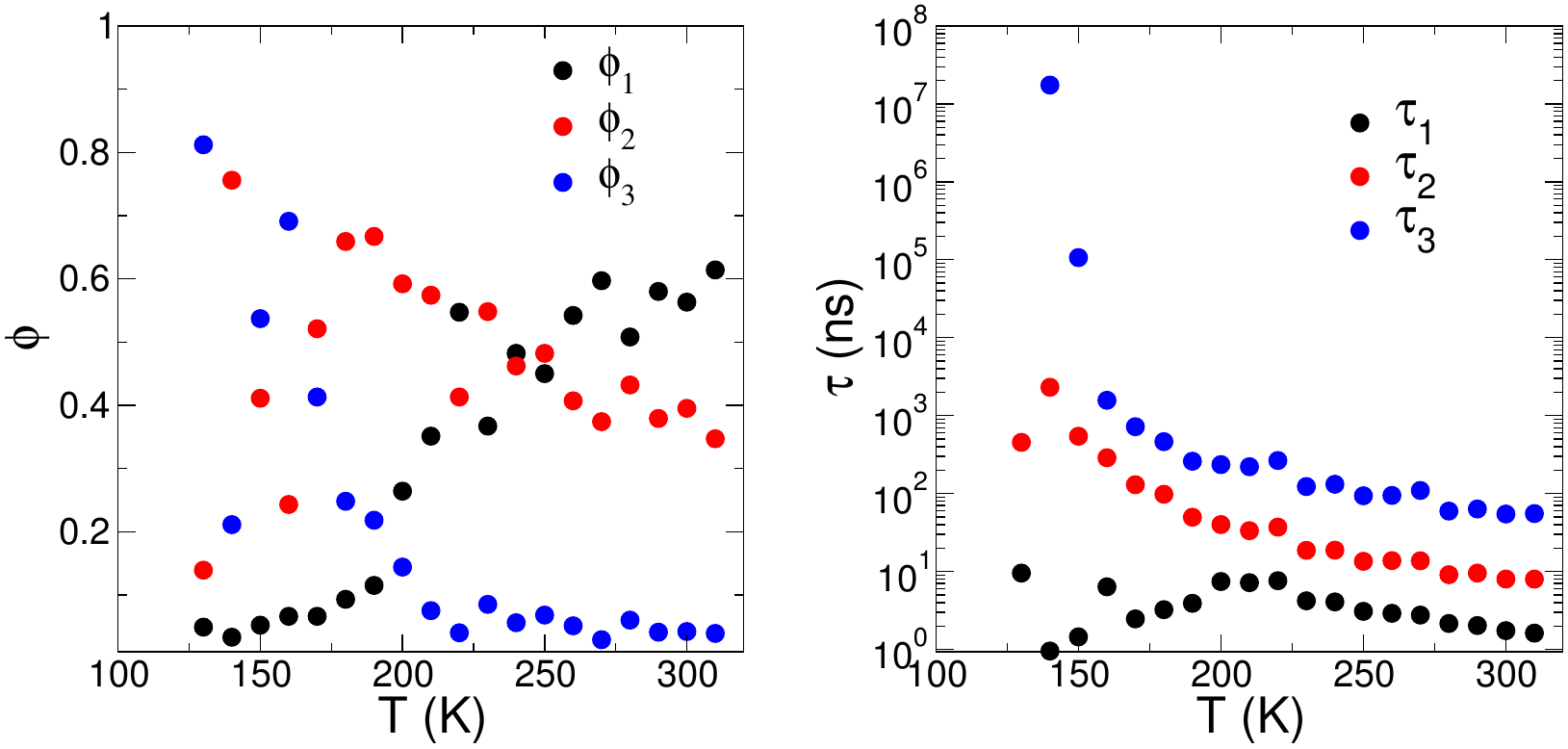}}
\caption{The parameters (weight ($\phi_i$) and time scale ($\tau_i$)) determined for the tri-exponential function used to fit the bulk water dynamics in the left panel of Fig.\ref{fig:water_preQ1}b. 
}\label{fig:weight_tau}
\end{center}
\end{figure}

\clearpage 
 


\begin{thebibliography}{10}

\bibitem{Dickerson1982}
Dickerson, R.~E.; Drew, H.~R.; Conner, B.~N.; Wing, R.~M.; Fratini, A.~V.;
  Kopka, M.~L.;
\newblock The anatomy of {A}-, {B}-, and {Z-DNA;}
\newblock {\em Science} {\bf 1982}, {\em 216}, 475--485.

\bibitem{Doster1989}
Doster, W.; Cusack, S.; Petry, W.;
\newblock Dynamical transition of myoglobin revealed by inelastic neutron
  scattering;
\newblock {\em Nature} {\bf 1989}, {\em 337}, 754--756.

\bibitem{Rupley1991}
Rupley, J.~A.; Careri, G.;
\newblock Protein hydration and function;
\newblock {\em Adv. Protein Chem.} {\bf 1991}, {\em 41}, 37--172.

\bibitem{Daniel2003}
Daniel, R.~M.; Dunn, R.~V.; Finney, J.~L.; Smith, J.~C.;
\newblock The role of dynamics in enzyme activity;
\newblock {\em Annu. Rev. Biophys. biomol. Struct.} {\bf 2003}, {\em 32},
  69--92.

\bibitem{Roh2006}
Roh, J.~H.; Curtis, J.~E.; Azzam, S.; Novikov, V.~N.; Peral, I.; Chowdhuri, X.;
  Gregory, R.~B.; Sokolov, A.~P.;
\newblock Influence of hydration on the dynamics of lysozyme;
\newblock {\em Biophys. J.} {\bf 2006}, {\em 91}, 2573--2588.

\bibitem{Caliskan2006}
Caliskan, G.; Briber, R.~M.; Thirumalai, D.; Garcia-Sakai, V.; Woodsen, S.~A.;
  Sokolov, A.~P.;
\newblock Dynamics transition in t{RNA} is solvent induced;
\newblock {\em J. Am. Chem. Soc.} {\bf 2006}, {\em 128}, 32--33.

\bibitem{Roh2009}
Roh, J.~H.; Briber, R.~M.; Damjanovic, A.; Thirumalai, D.; Woodsen, S.~A.;
  Sokolov, A.~P.;
\newblock {Dynamics of tRNA at different levels of hydration;}
\newblock {\em Biophys. J.} {\bf 2009}, {\em 96}, 2755--2762.

\bibitem{Fitter1996PNAS}
Fitter, J.; Lechner, R.; Buldt, G.; Dencher, N.;
\newblock Internal molecular motions of bacteriorhodopsin: hydration-induced
  flexibility studied by quasielastic incoherent neutron scattering using
  oriented purple membranes;
\newblock {\em Proc. Natl. Acad. Sci. U.S.A.} {\bf 1996}, {\em 93}(15),
  7600--7605.

\bibitem{Reat2000PNAS}
R{\'e}at, V.; Dunn, R.; Ferrand, M.; Finney, J.~L.; Daniel, R.~M.; Smith,
  J.~C.;
\newblock Solvent dependence of dynamic transitions in protein solutions;
\newblock {\em Proc. Natl. Acad. Sci. U.S.A.} {\bf 2000}, {\em 97}(18),
  9961--9966.

\bibitem{Tsai00BJ}
Tsai, A.~M.; A.Neumann, D.; Bell, L.~N.;
\newblock {Molecular Dynamics of Solid-State Lysozyme as Affected by Glycerol
  and Water: A Neutron Scattering Study;}
\newblock {\em Biophys. J.} {\bf 2000}, {\em 79}, 2728--2732.

\bibitem{Ramunssen92Nature}
Rasmussen, B.~F.; Stock, A.~M.; Ringe, D.; Petsko, G.~A.;
\newblock {Crystalline ribonuclease A loses function below the dynamical
  transition at 220 K;}
\newblock {\em Nature} {\bf 1992}, {\em 357}, 423 -- 424.

\bibitem{tao1989BP}
Tao, N.; Lindsay, S.; Rupprecht, A.;
\newblock Structure of dna hydration shells studied by raman spectroscopy;
\newblock {\em Biopolymers} {\bf 1989}, {\em 28}(5), 1019--1030.

\bibitem{denisov1997JMB}
Denisov, V.~P.; Carlstr{\"o}m, G.; Venu, K.; Halle, B.;
\newblock {Kinetics of DNA hydration;}
\newblock {\em J. Mol. Biol.} {\bf 1997}, {\em 268}(1), 118--136.

\bibitem{Endres2004}
Endres, R.~G.; Cox, D.~L.; Singh, R. R.~P.;
\newblock Colloquium: The quest for high-conductance {DNA;}
\newblock {\em Rev. Mod. Phys.} {\bf 2004}, {\em 76}, 195--214.

\bibitem{Egli96Biochem}
Egli, M.; Portmann, S.; Usman, N.;
\newblock {RNA hydration: a detailed look;}
\newblock {\em Biochemistry} {\bf 1996}, {\em 35}(26), 8489--8494.

\bibitem{Auffinger1997}
Auffinger, P.; Westhof, E.;
\newblock {RNA} hydration: Three nanoseconds of multiple molecular dynamics
  simulations of the solvated t{RNA} anticoden hairpin;
\newblock {\em J. Mol. Biol.} {\bf 1997}, {\em 269}, 326--341.

\bibitem{Chen08PRE}
Chu, X.-Q.; Fratini, E.; Baglioni, P.; Faraone, A.; Chen, S.-H.;
\newblock Observation of a dynamic crossover in rna hydration water which
  triggers a dynamic transition in the biopolymer;
\newblock {\em Phys. Rev. E.} {\bf 2008}, {\em 77}(1), 011908.

\bibitem{Chu13JPCL}
Chu, X.-Q.; Mamontov, E.; O'Neill, H.; Zhang, Q.;
\newblock Temperature dependence of logarithmic-like relaxational dynamics of
  hydrated trna;
\newblock {\em J. Phys. Chem. Lett.} {\bf 2013}, {\em 4}(6), 936--942.

\bibitem{Zhang13JPCL}
Zhang, H.; Khodadadi, S.; Fiedler, S.~L.; Curtis, J.~E.;
\newblock {Role of Water and Ions on the Dynamical Transition of RNA;}
\newblock {\em J. Phys. Chem. Lett.} {\bf 2013}, {\em 4}(19), 3325--3329.

\bibitem{Rinnenthal10NAR}
Rinnenthal, J.; Klinkert, B.; Narberhaus, F.; Schwalbe, H.;
\newblock {Direct observation of the temperature-induced melting process of the
  Salmonella fourU RNA thermometer at base-pair resolution;}
\newblock {\em Nucleic Acids Research} {\bf 2010}, {\em 38}(11), 3834--3847.

\bibitem{Nikolova10RNA}
Nikolova, E.~N.; Al-Hashimi, H.~M.;
\newblock {Thermodynamics of RNA melting, one base pair at a time;}
\newblock {\em RNA} {\bf 2010}, {\em 16}(9), 1687--1691.

\bibitem{Zhang06Science}
Zhang, Q.; Sun, X.; Watt, E.~D.; Al-Hashimi, H.~M.;
\newblock Resolving the motional modes that code for rna adaptation;
\newblock {\em Science} {\bf 2006}, {\em 311}(5761), 653--656.

\bibitem{Yoon13JACS}
Yoon, J.; Thirumalai, D.; Hyeon, C.;
\newblock {Urea-induced denaturation of preQ1-riboswitch;}
\newblock {\em J. Am. Chem. Soc.} {\bf 2013}, {\em 135}, 12112--12121.

\bibitem{Kuhrova14JCTC}
K{\"u}hrov{\'a}, P.; Otyepka, M.; Sponer, J.; Banas, P.;
\newblock {Are Waters around RNA More than Just a Solvent?--An Insight from
  Molecular Dynamics Simulations;}
\newblock {\em J. Chem. Theor. Comp.} {\bf 2013}, {\em 10}(1), 401--411.

\bibitem{WoodsonJMBII01}
Heilman-Miller, S.~L.; Pan, J.; Thirumalai, D.; Woodson, S.~A.;
\newblock {Role of Counterion Condensation in Folding of \emph{Tetrahymena}
  Ribozyme II. Counterion-dependence of Folding Kinetics;}
\newblock {\em J. Mol. Biol.} {\bf 2001}, {\em 309}, 57--68.

\bibitem{MisraPNAS01}
Misra, V.~K.; Draper, D.~E.;
\newblock A thermodynamic framework for {Mg}$^{2+}$ binding to {RNA;}
\newblock {\em Proc. Natl. Acad. Sci. U.S.A.} {\bf 2001}, {\em 98}(22),
  12456--12461.

\bibitem{Chen09JMB}
Chen, A.~A.; Draper, D.~E.; Pappu, R.~V.;
\newblock {Molecular simulation studies of monovalent counterion-mediated
  interactions in a model RNA kissing loop;}
\newblock {\em J. Mol. Biol.} {\bf 2009}, {\em 390}(4), 805--819.

\bibitem{kirmizialtin10JPCB}
Kirmizialtin, S.; Elber, R.;
\newblock {Computational exploration of mobile ion distributions around RNA
  duplex;}
\newblock {\em J. Phys. Chem. B} {\bf 2010}, {\em 114}(24), 8207--8220.

\bibitem{Kirmizialtin12BJ}
Kirmizialtin, S.; Pabit, S.~A.; Meisburger, S.~P.; Pollack, L.; Elber, R.;
\newblock {RNA and its ionic cloud: solution scattering experiments and
  atomically detailed simulations;}
\newblock {\em Biophys. J.} {\bf 2012}, {\em 102}(4), 819--828.

\bibitem{Hayes12JACS}
Hayes, R.~L.; Noel, J.~K.; Mohanty, U.; Whitford, P.~C.; Hennelly, S.~P.;
  Onuchic, J.~N.; Sanbonmatsu, K.~Y.;
\newblock {Magnesium fluctuations modulate RNA dynamics in the SAM-I
  riboswitch;}
\newblock {\em J. Am. Chem. Soc.} {\bf 2012}, {\em 134}(29), 12043--12053.

\bibitem{Krasovska06BJ}
Krasovska, M.~V.; Sefcikova, J.; R{\'e}blov{\'a}, K.; Schneider, B.; Walter,
  N.~G.; {\v{S}}poner, J.;
\newblock {Cations and hydration in catalytic RNA: molecular dynamics of the
  hepatitis delta virus ribozyme;}
\newblock {\em Biophys. J.} {\bf 2006}, {\em 91}(2), 626--638.

\bibitem{Mandal2003Cell}
Mandal, M.; Boes, B.; Barrick, J.~E.; Winkler, W.~C.; Breaker, R.~R.;
\newblock Riboswitches control fundamental biochemical pathways in {B}acillus
  subtilis and other bacteria;
\newblock {\em Cell} {\bf 2003}, {\em 113}, 577--586.

\bibitem{Mandal2003NSMB}
Mandal, M.; Breaker, R.~R.;
\newblock Adenine riboswitches and gene activation by disruption of a
  transcription terminator;
\newblock {\em Nat. Struct. Mol. Biol.} {\bf 2003}, {\em 11}, 29--35.

\bibitem{Serganov2004}
Serganov, A.; Yuan, Y.~T.; Pikovskaya, O.; Polonskaia, A.; Malinina, L.; Phan,
  A.~T.; Hobartner, C.; Micura, R.; Breaker, R.~R.; Patel, D.;
\newblock Structural basis for discriminative regulation of gene expression by
  adenine- and guanine-sensing m{RNA}s;
\newblock {\em J. Chem. Biol.} {\bf 2004}, {\em 11}, 1729--1741.

\bibitem{Winkler2005}
Winkler, W.~C.; Breaker, R.~R.;
\newblock Regulation of bacterial gene expression by riboswitches;
\newblock {\em Annu. Rev. Microbiol.} {\bf 2005}, {\em 59}, 487--517.

\bibitem{Cheah2007}
Cheah, M.~T.; Wachter, A.; Sudarsan, N.; Breaker, R.~R.;
\newblock Control of alternative {RNA} splicing and gene expression by
  eukaryotic riboswitches;
\newblock {\em Nature} {\bf 2007}, {\em 447}, 497--500.

\bibitem{Montange2008}
Montange, R.~K.; Batey, R.~T.;
\newblock Riboswitches: emerging themes in {RNA} structure and function;
\newblock {\em Annu. Rev. Biophys.} {\bf 2008}, {\em 37}, 117--133.

\bibitem{Zhang11JACS}
Zhang, Q.; Kang, M.; Peterson, R.~D.; Feigon, J.;
\newblock {Comparison of Solution and Crystal Structures of PreQ$_1$ Riboswitch
  Reveals Calcium-Induced Changes in Conformation and Dynamics;}
\newblock {\em J. Am. Chem. Soc.} {\bf 2011}, {\em 133}, 5190--5193.

\bibitem{Kumar07JCP}
Kumar, R.; Schmidt, J.~R.; Skinner, J.~L.;
\newblock {Hydrogen bonding definitions and dynamics in liquid water;}
\newblock {\em J. Chem. Phys.} {\bf 2007}, {\em 126}, 204107.

\bibitem{Williams94ProtSci}
Williams, M.~A.; Goodfellow, J.~M.; Thornton, J.~M.;
\newblock {Buried waters and internal cavities in monomeric proteins;}
\newblock {\em Protein Sci.} {\bf 1994}, {\em 3}, 1224--1235.

\bibitem{Tarek00BJ}
Tarek, M.; Tobias, D.~J.;
\newblock {The Dynamics of Protein Hydration Water: A Quantitative Comparison
  of Molecular Dynamics Simulations and Neutron-scattering Experiments;}
\newblock {\em Biophys. J.} {\bf 2000}, {\em 79}, 3244--3257.

\bibitem{denisov1997JPCB}
Denisov, V.~P.; Venu, K.; Peters, J.; H{\"o}rlein, H.~D.; Halle, B.;
\newblock {Orientational disorder and entropy of water in protein cavities;}
\newblock {\em J. Phys. Chem. B.} {\bf 1997}, {\em 101}(45), 9380--9389.

\bibitem{oostenbrink04JCC}
Oostenbrink, C.; Villa, A.; Mark, A.; Van~Gunsteren, W.;
\newblock {A biomolecular force field based on the free enthalpy of hydration
  and solvation: The GROMOS force-field parameter sets 53A5 and 53A6;}
\newblock {\em J. Comp. Chem.} {\bf 2004}, {\em 25}(13), 1656--1676.

\bibitem{Fenimore02PNAS}
Fenimore, P.~W.; Frauenfelder, H.; McMahon, B.~H.; Parak, F.~G.;
\newblock Slaving: solvent fluctuations dominate protein dynamics and
  functions;
\newblock {\em Proc. Natl. Acad. Sci. U.S.A.} {\bf 2002}, {\em 99}(25),
  16047--16051.

\bibitem{Thirum05Biochem}
Thirumalai, D.; Hyeon, C.;
\newblock {RNA and Protein folding: Common Themes and Variations;}
\newblock {\em Biochemistry} {\bf 2005}, {\em 44}(13), 4957--4970.

\bibitem{Auffinger02BiophysChem}
Auffinger, P.; Westhof, E.;
\newblock {Melting of the solvent structure around a RNA duplex: a molecular
  dynamics simulation study;}
\newblock {\em Biophys. Chem.} {\bf 2002}, {\em 95}(3), 203--210.

\bibitem{king2012JACS}
King, J.~T.; Kubarych, K.~J.;
\newblock {Site-specific coupling of hydration water and protein flexibility
  studied in solution with ultrafast 2D-IR spectroscopy;}
\newblock {\em J. Am. Chem. Soc.} {\bf 2012}, {\em 134}(45), 18705--18712.

\bibitem{Khodadadi2011JPCB}
Khodadadi, S.; Curtis, J.; Sokolov, A.~P.;
\newblock {Nanosecond relaxation dynamics of hydrated proteins: Water versus
  protein contributions;}
\newblock {\em J. Phys. Chem. B.} {\bf 2011}, {\em 115}(19), 6222--6226.

\bibitem{lee2001Nature}
Lee, A.~L.; Wand, A.~J.;
\newblock Microscopic origins of entropy, heat capacity and the glass
  transition in proteins;
\newblock {\em Nature} {\bf 2001}, {\em 411}(6836), 501--504.

\bibitem{Gallo1996PRL}
Gallo, P.; Sciortino, F.; Tartaglia, P.; Chen, S.-H.;
\newblock Slow dynamics of water molecules in supercooled states;
\newblock {\em Phys. Rev. Lett.} {\bf 1996}, {\em 76}(15), 2730.

\bibitem{EyringJCP35}
Eyring, H.;
\newblock The activated complex in chemical reactions;
\newblock {\em J. Chem. Phys.} {\bf 1935}, {\em 3}, 107--115.

\bibitem{Tournier03BiophysJ}
Tournier, A.; Xu, J.; Smith, J.;
\newblock {Translational hydration water dynamics drives the protein glass
  transition;}
\newblock {\em Biophys. J.} {\bf 2003}, {\em 85}, 1871--1875.

\bibitem{Luzar96PRL}
Luzar, A.; Chandler, D.;
\newblock Effect of environment on hydrogen bond dynamics in liquid water;
\newblock {\em Phys. Rev. Lett.} {\bf 1996}, {\em 76}, 928--931.

\bibitem{Allner2012JCTC}
Alln{\'e}r, O.; Nilsson, L.; Villa, A.;
\newblock Magnesium ion--water coordination and exchange in biomolecular
  simulations;
\newblock {\em J. Chem. Theory Compt.} {\bf 2012}, {\em 8}(4), 1493--1502.

\bibitem{Caliskan05PRL}
Caliskan, G.; Hyeon, C.; Perez-Salas, U.; Briber, R.~M.; Woodson, S.~A.;
  Thirumalai, D.;
\newblock {Persistence Length Changes Dramatically as RNA Folds;}
\newblock {\em Phys. Rev. Lett.} {\bf 2005}, {\em 95}, 268303.

\bibitem{Roh11JACS}
Roh, J.~H.; Tyagi, M.; Briber, R.~M.; Woodson, S.~A.; Sokolov, A.~P.;
\newblock {The Dynamics of Unfolded versus Folded tRNA: The Role of
  Electrostatic Interactions;}
\newblock {\em J. Am. Chem. Soc.} {\bf 2011}, {\em 133}, 16406--16409.

\bibitem{AlHashimi08COSB}
Al-Hashimi, H.; Walter, N.;
\newblock {RNA dynamics: it is about time;}
\newblock {\em Curr. Opin. Struct. Biol.} {\bf 2008}, {\em 18}(3), 321--329.

\bibitem{Solomatin10Nature}
Solomatin, S.~V.; Greenfeld, M.; Chu, S.; Herschlag, D.;
\newblock Multiple native states reveal persistent ruggedness of an {RNA}
  folding landscape;
\newblock {\em Nature} {\bf 2010}, {\em 463}, 681--684.

\bibitem{Hyeon2012NatureChem}
Hyeon, C.; Lee, J.; Yoon, J.; Hohng, S.; Thirumalai, D.;
\newblock Hidden complexity in the isomerization dynamics of holliday
  junctions;
\newblock {\em Nature Chem.} {\bf 2012}, {\em 4}, 907--914.

\bibitem{Herskovits63Biochemistry}
Herskovits, T.~T.;
\newblock {Nonaqueous Solutions of DNA; Denaturation by Urea and Its Methyl
  Derivatives;}
\newblock {\em Biochemistry} {\bf 1963}, {\em 2}(2), 335--340.

\bibitem{Denning13BiophysChem}
Denning, E.~J.; Thirumalai, D.; MacKerell~Jr, A.~D.;
\newblock {Protonation of trimethylamine N-oxide (TMAO) is required for
  stabilization of RNA tertiary structure;}
\newblock {\em Biophys. Chem.} {\bf 2013}, {\em 184}, 8--16.

\bibitem{Philips2005}
Philips, J.~C.; Braun, R.; Wang, W.; Gumbart, J.; Tajkhorshid, E.; Villa, E.;
  Chipot, C.; Skeel, R.~D.; Kale, L.; Schulten, K.;
\newblock Scalable molecular dynamics with {NAMD;}
\newblock {\em J. Comp. Chem.} {\bf 2005}, {\em 26}, 1781--1802.

\bibitem{Humphrey1996}
Humphrey, W.; Dalke, A.; Schulten, K.;
\newblock {VMD: visual molecular dynamics;}
\newblock {\em J. Mol. Graphics} {\bf 1996}, {\em 14}, 33--38.

\bibitem{Darden1993}
Darden, T.; York, D.; Pedersen, L.;
\newblock {Particle mesh Ewald: An $N\cdot\log{N}$ method for Ewald sums in
  large systems;}
\newblock {\em J. Chem. Phys.} {\bf 1993}, {\em 98}(12), 10089--10092.

\bibitem{Ryckaert1977}
Ryckaert, J.-P.; Ciccotti, G.; Berendsen, H.~J.;
\newblock Numerical integration of the cartesian equations of motion of a
  system with constraints: molecular dynamics of $n$-alkanes;
\newblock {\em J. Comp. Phys.} {\bf 1977}, {\em 23}(3), 327--341.

\bibitem{Chen13PNAS}
Chen, A.~A.; Garc{\'\i}a, A.~E.;
\newblock {High-resolution reversible folding of hyperstable RNA tetraloops
  using molecular dynamics simulations;}
\newblock {\em Proc. Natl. Acad. Sci. U.S.A.} {\bf 2013}, {\em 110}(42),
  16820--16825.

\bibitem{Chandler1996}
Luzar, A.; Chandler, D.;
\newblock Hydrogen-bond kinetics in liquid water;
\newblock {\em Nature} {\bf 1996}, {\em 379}, 55--57.

\bibitem{Baker01PNAS}
Baker, N.~A.; Sept, D.; Joseph, S.; Holst, M.~H.; {McCammon}, J.~A.;
\newblock {Electrostatics of Nanosystems: Application to microtubules and the
  ribosome;}
\newblock {\em Proc. Natl. Acad. Sci. U.S.A.} {\bf 2001}, {\em 102},
  13070--13074.

\end{thebibliography}

\clearpage

\end{document}